# Unconventional Optical Rotation in the Charge Ordered State of Kagome Metal CsV$_3$Sb$_5$

Camron Farhang[1], Jingyuan Wang[1], Brenden R. Ortiz[2], Stephen D. Wilson[2] and Jing Xia[1*]

[1]Department of Physics and Astronomy, University of California, Irvine, California 92697, USA
[2]Materials Department, University of California, Santa Barbara, Santa Barbara, CA 93106, USA.11973

**Kagome metals AV$_3$Sb$_5$ (A = K, Cs, Rb) provide a rich platform for intertwined orders, where evidence for time-reversal symmetry breaking, likely due to the long-sought loop currents, has emerged in STM and muon spin relaxation experiments. An isotropic component in the spontaneous optical rotation has also been reported, and it was interpreted as the magneto-optic Kerr effect arising from time-reversal symmetry breaking. Intriguingly, the observed rotations differ by five orders of magnitude between different wavelengths and samples, suggesting more intricate physics. Here we report optical rotation and polar Kerr measurements in CsV$_3$Sb$_5$ crystals at the same wavelength. We observe large isotropic components of 1 milliradian in the optical rotation that do not respond to applied magnetic fields but flip sign between samples, while the spontaneous Kerr signal is less than 20 nanoradian. Our results prove unambiguously that the reported isotropic rotation is not related to time-reversal symmetry breaking but instead indicates a new intertwined order that onsets at the same temperature.**

*Email: xia.jing@uci.edu

The Kagome lattice is a rich platform for novel phases of matter due to the interplay between strong correlation and topological orders, such as in the case of Chern topological magnet TbMn$_6$Sn$_6$ [1]. In particular the recently discovered quasi-two-dimensional Kagome compounds AV$_3$Sb$_5$ (A=K, Rb and Cs) [2,3] have elaborate phase diagrams due to the ideal Kagome network governed by layers of vanadium and antimony intercalated by alkali metal ions [2,3]. The fascinating electronic band structure containing Dirac cones, flat bands, and Van Hove singularities [3,4] leads to intertwined [5] electronic instabilities and exhibits charge density wave (CDW) [6–10], pressure-tunable superconductivity[3,11–14], and time reversal symmetry breaking (TRSB) that has been revealed by STM [15] and muon-spin relaxation (μSR) [16–18] experiments. The TRSB state may be related to the long-sought loop currents[19,20], and could produce a spontaneous magneto-optic Kerr effect (MOKE) [21,22] rotation $\theta_K$, which arises from the optical phase difference $\Delta\varphi = 2\theta_K$ between counter-propagating circularly polarized light beams that are time-reversal images of each other.

Indeed, two optical rotation experiments performed both at 800 $nm$ wavelength have revealed [23,24] a spontaneous total rotation $\theta_T$ in the form of $\theta_T = \theta_C + \theta_P \sin(2\alpha - A)$, where $\alpha$ is the incident polarization angle, and $A$ is the principle axis direction. $\theta_P$ represents the anisotropic component due to the nematicity [23,24] of the CDW state. Since translational symmetry is also broken, it will be called "anisotropy" in this work to avoid confusions. $\theta_C$ represents the isotropic (polarization independent) component and was found to be 0.5 $mrad$ [24] and 50 $\mu rad$ [23] in CsV$_3$Sb$_5$. In transmission, such an isotropic $\theta_C$ component could arise from optical activities in a chiral material, which causes an optical phase difference between circularly polarized lights propagating in the same direction. However, this effect is usually rejected [25] in normal incidence reflection used in the above experiments. Therefore, the observed isotropic rotation $\theta_C$ has been naturally attributed to a spontaneous MOKE signal $\theta_K$ due to TRSB at $T_{CDW}$ [23,24]. Intriguingly, it is noted [24] that such a large $\theta_C$ can be compared with that of some magnetic materials, while both μSR [16–18] experiments and theoretical calculations [26] indicate extremely small sub-gauss level magnetic flux density that would usually lead to nano-radians to sub-microradian levels of spontaneous MOKE signals [27,28]. Dedicated MOKE experiments [29–31] have been carried out in CsV$_3$Sb$_5$ at 1550 $nm$ wavelength utilizing zero-loop Sagnac interferometers [32] in an attempt to resolve this puzzle. Such an interferometer is sensitive only to TRSB effects by detecting the optical phase difference $\Delta\varphi$ between time-reversed counter-propagating circularly polarized light beams, which is twice the MOKE angle: $\Delta\varphi = 2\theta_K$. Surprisingly the 1550 $nm$ MOKE experiments report either a much smaller spontaneous Kerr signal $\theta_K \sim 2\ \mu rad$ [30] or near-zero values $\theta_K < 0.03\ \mu rad$ [29,31]. To explain this giant discrepancy, one is tempted to assume a resonance enhancement at 800 $nm$ wavelength over 1550 $nm$. On the contrary the near-infrared spectra of CsV$_3$Sb$_5$ are rather flat [33,34] with a Lorentz resonance at 6000 $cm^{-1}$ (equivalent to 1667 $nm$) [34] suggesting instead a larger expected signal at the 1550 $nm$ wavelength.

To solve this outstanding mystery, we perform optical rotation ($\theta_T$) and polar MOKE ($\theta_K$) measurements at 1550 $nm$ wavelength on the same CsV$_3$Sb$_5$ crystals. Rather surprisingly, we observe giant $\theta_C \sim \pm 1200\ \mu rad$ below $T_{CDW}$ but negligible $\theta_K < 0.02\ \mu rad$ at zero magnetic fields. The $\theta_C$ component doesn't respond to applied magnetic fields, which is opposite to MOKE and to the magnetic field responses found in STM [15] and μSR [16–18] experiments. Therefore, we conclude that the observed $\theta_C$ represents an unconventional optical rotation that is not due to either TRSB, anisotropy or chiral order, but originates from a new intertwined order that onsets at $T_{CDW}$.

High quality CsV$_3$Sb$_5$ single crystals, dubbed sample 1 and 2, were grown by the self-flux method at UCSB[3]. The first order CDW transition at $T_{CDW} \sim 94\ K$ is clearly characterized by sharp peaks in the heat capacity $C_P$ and resistance derivative $dR/dT$ as shown in Fig.1(e). The crystals were cleaved perpendicular to the c-axis to expose optically flat areas and were mounted to the sample stage using Ge-varnish for minimal strain.

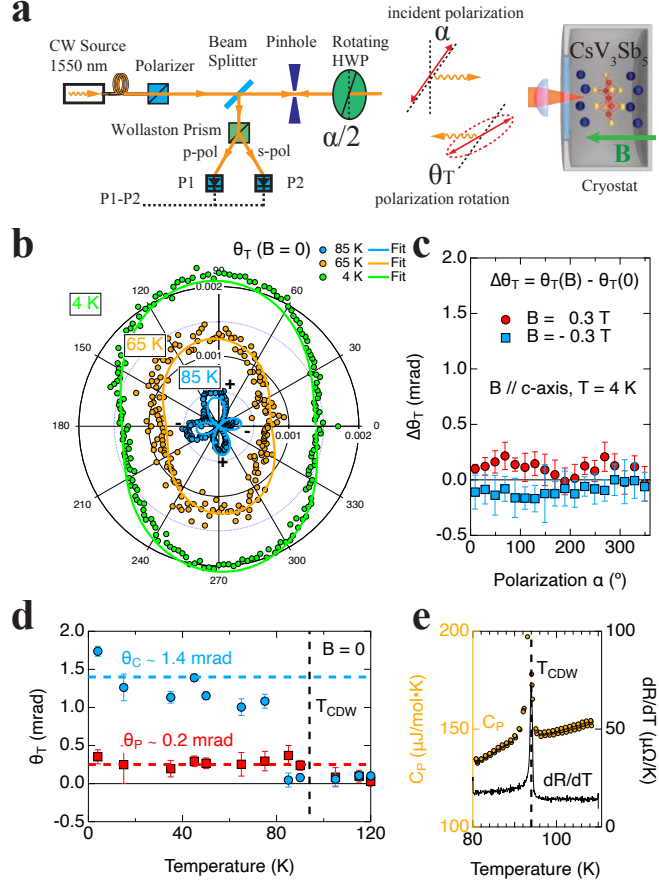

**Fig. 1. Optical rotation $\theta_T$ of CsV$_3$Sb$_5$ sample 1.** (a) Polarization rotation $\theta_T$ at normal incidence is measured as a function of incident polarization angle $\alpha$. See SI for details. (b) Polar plots of spontaneous $\theta_T$ at representative temperatures, fitted with $\theta_T = \theta_C + \theta_P \sin(2\alpha - A)$. See SI for full data. (c) Changes of rotation $\Delta\theta_T$ by $\pm 0.3\ T$ magnetic fields at $4\ K$, showing opposite magnetic fields fail to flip the sign of $\theta_C$. (d) Fitted $\theta_C$ ($\theta_P$) up to $1.4\ mrad$ ($0.2\ mrad$) with onsets below $T_{CDW}$. (e) $T_{CDW} \sim 94\ K$ is marked by sharp peaks in the specific heat $C_P$ and resistance derivative $dR/dT$.

As shown in Fig.1(a), a standard optical setup based on a Wollaston prism is used to measure the optical rotation ($\theta_T$) as a function of incident polarization angle $\alpha$. And a zero-loop Sagnac interferometer [32] as shown in Fig.2(a) is used for polar MOKE ($\theta_K$) detection and imaging. Both instruments are connected to the same optical cryostat so $\theta_T$ and $\theta_K$ can be obtained from the same region in the sample during one experiment. Operation and calibration of both instruments on several test samples are described in the Supplementary Information. As shown in Fig.S2, the resolution of $\theta_T$ and $\theta_K$ are $30\ \mu rad$ and $0.02\ \mu rad$ respectively. And they agree within 2% on the MOKE signal of a magnetic film test sample. The root cause for the Sagnac interferometer's superior sensitivity is that it only measures microscopic TRSB and rejects any non-TRSB effects such as anisotropy. This is because the sourcing aperture for one light is the receiving aperture for the other time-reversed counterpropagating light, both apertures being the same single mode optical fiber. Hence Onsager's relations guarantee zero signal in the absence of microscopic TRSB. Such rejection of non-TRSB effect is demonstrated in Fig.S2(e) to $0.04\ \mu rad$ level with an anisotropic polymer film, which displays anisotropic polarization rotations with 2-fold rotational symmetry of $\pm 20\ mrad$.

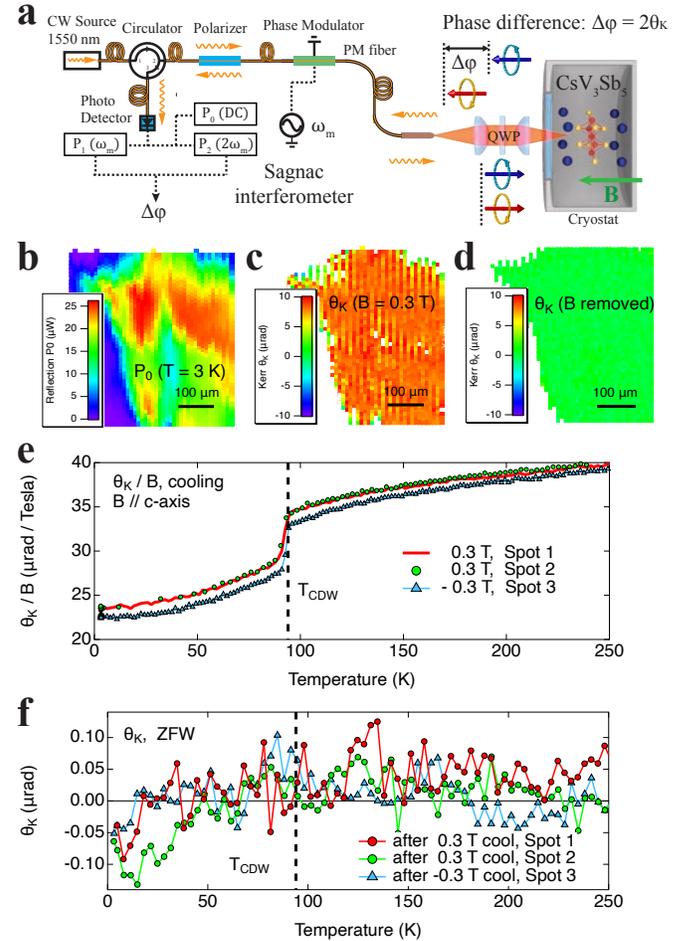

**Fig. 2. MOKE $\theta_K$ in CsV$_3$Sb$_5$ sample 1.** (a) MOKE signal $\theta_K$ is measured by a Sagnac interferometer during the same experiment. See SI for details. (b)(c) Scanning images of reflected optical power $P_0$ and MOKE $\theta_K$ at $3\ K$ in $B = 0.3\ T$ showing uniform $\theta_K$ in the reflective regions. (d) Image of $\theta_K$ after removal of the magnetic field, showing zero spontaneous MOKE signal across the sample. (e) $\theta_K/B$ during cooldown at a few locations and magnetic fields showing a paramagnetic MOKE response that drops sharply below $T_{CDW}$. (f) No onset of spontaneous $\theta_K$ at $T_{CDW}$ was found during subsequent zero field warmup (ZFW).

We first examine CsV$_3$Sb$_5$ sample 1, whose optical rotation and MOKE results are summarized in Fig.1 and Fig.2 respectively. At $B = 0$, polar plots of $\theta_T(\alpha)$ at representative temperatures are shown in Fig.1(b), with additional temperatures plotted in Fig.S3(a)-(l). Below $T_{CDW}$, the total rotation $\theta_T$ contains an anisotropic component $\theta_P$ and an isotropic component $\theta_C$, similar to the findings at the 800 nm wavelength [23,24]. Fitted values of $\theta_P$ and $\theta_C$ are plotted in Fig.1(d) showing sharp onsets just below $T_{CDW}$. The $\theta_P$ component originates from the reduction from 6-fold rotational symmetry to 2-fold in the CDW state. Since electronic nematicity was found [35,36] in CsV$_3$Sb$_5$ at much lower temperatures, $\theta_P$ is likely of structural origin. The size of $\theta_P$ is 0.2 $mrad$ in our 1550 $nm$ measurement. It should scale linearly with the anisotropies in the permittivity and magneto-electric tensors [37] and inversely with the wavelength $\lambda$. The rather flat near-infrared spectra [33,34] of CsV$_3$Sb$_5$ suggest that the former factor is comparable between 800 $nm$ and 1550 $nm$. Hence, $\theta_P$ is expected to roughly double at 800 $nm$ reaching 0.4 $mrad$, which indeed falls between the experimentally reported values of 0.9 $mrad$ [24] and 0.2 $mrad$ [23] at 800 $nm$.

What comes as a surprise is the size of the isotropic component $\theta_C$, which has been interpreted as MOKE ($\theta_K$) in the 800 $nm$ experiments [23,24] and is expected to be vanishingly small at 1550 $nm$. Instead, the observed $\theta_C$ as plotted in Fig.1(d) reaches 1.4 $mrad$, which is *larger* than the reported values [23,24] at 800 $nm$. A MOKE signal, which arises from TRSB, should flip sign with opposite magnetic fields. In contrast, we found no such response in $\theta_C$: as shown in Fig.1(c) with $\pm 0.3\ T$ applied magnetic fields, the change of optical rotation $\Delta\theta_T$ is much smaller than $\theta_C$, approaching the noise level. This field-insensitivity is also opposite to the TRSB signatures reported in both STM [15] and μSR [16–18] experiments, which show clear magnetic responses. All these observations are suggestive that the isotropic component $\theta_C$ reported in [23,24] and in this work is not MOKE and is not related to TRSB.

The decisive evidence that $\theta_C$ is not MOKE comes from Sagnac measurements of the same sample. As explained earlier, a Sagnac interferometer is only sensitive to MOKE, which is a direct result of TRSB. Fig.2(c) and (d) are $\theta_K$ images of the same region in sample 1 at 3 $K$ with $B = 0.3\ T$ and $B = 0$ respectively. In the optically flat region where the reflected optical power (Fig.2(b)) $P_0 > 1\ \mu W$, the MOKE signal is uniform with $\theta_K \sim 7\ \mu rad$ in $B = 0.3\ T$, and vanishes after the field is removed. $\theta_K$ is found to be paramagnetic and the MOKE susceptibility $\theta_K/B$ remains field-independent during cooldowns as shown in Fig.2(e). In another study [31] we verified this paramagnetic MOKE response at fields up to 9 $T$. The absence of a Curie–Weiss shape in $\theta_K/B$ leads us to attribute it to the Pauli paramagnetism. And the sharp drop in $\theta_K/B$ from 32 $\mu rad/T$ to 27 $\mu rad/T$ at $T_{CDW}$ is likely due to a decreased density of states in the CDW phase, which agrees with the reported reduction of magnetic susceptibility [3]. The magnetic fields were removed at the base temperature, and the subsequent zero field warmups (ZFW) are plotted in Fig.2(f). We observe no onset of spontaneous $\theta_K$ at $T_{CDW}$ with an uncertainty of 0.02 $\mu rad$, which is five orders of magnitude smaller than the observed $\theta_C \sim 1.4\ mrad$ in the same sample.

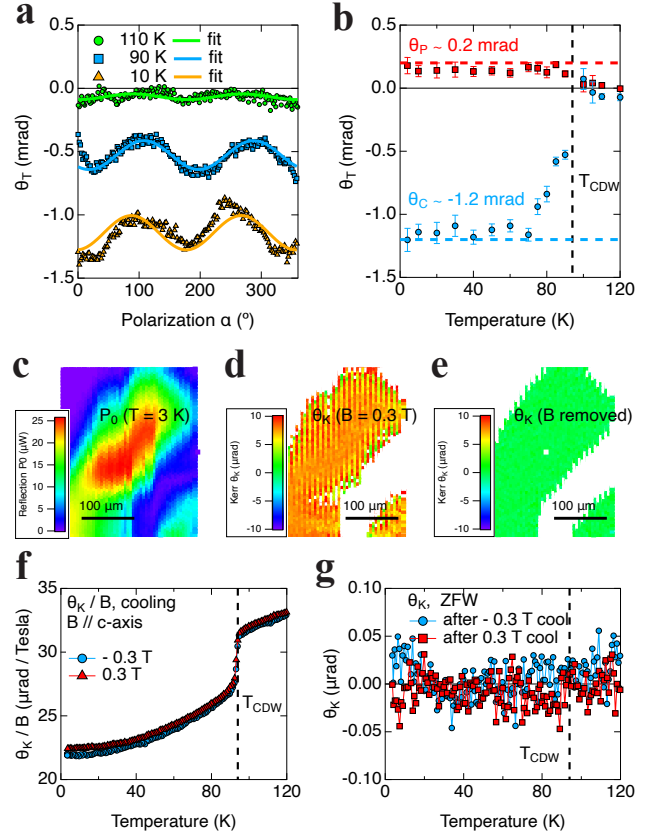

Fig. 3. Optical rotation and MOKE in another CsV$_3$Sb$_5$ sample 2 with negative $\theta_C$. (a) Spontaneous $\theta_T$ at representative temperatures, fitted with $\theta_T = \theta_C + \theta_P \sin(2\alpha - A)$. See SI for full data. (b) Fitted $\theta_C$ ($\theta_P$) up to $-1.2\ mrad$ (0.2 $mrad$) with onsets below $T_{CDW}$. Note the negative sign of $\theta_C$. (c), (d) and (e) are scanning images of $P_0$, $\theta_K$ with $B = 0.3\ T$, and $\theta_K$ after field removal, all at 3 $K$. (f) and (g) are averaged traces of $\theta_K/B$ during field cooldowns and of spontaneous $\theta_K$ during subsequent ZFW. See SI for full data. There is no observable onset of spontaneous $\theta_K$ at $T_{CDW}$ with an uncertainty of 0.02 $\mu rad$.

The sign of $\theta_C$ is also unrelated to that of $\theta_K$, and appears to have already been determined during crystal growth. On the same crystal, we have always found the same sign of $\theta_C$, regardless of direction of applied magnetic fields. The sign of $\theta_C$ can flip between samples. One such example can be found in CsV$_3$Sb$_5$ sample 2. The polarization rotations $\theta_T(\alpha)$ of sample 2 are mostly negative, as shown for representative temperatures in Fig.3(a), and for additional temperatures in Fig.S4(a)-(o). The fitted anisotropic component $\theta_P$ and isotropic component $\theta_C$ are plotted in Fig.3(b), both showing sharp onsets at $T_{CDW}$. The low-temperature value of $\theta_P$ (0.2 $mrad$) is comparable to that in sample 1, while $\theta_C \sim -1.2\ mrad$ is opposite to that found in sample 1. In one 800 $nm$ experiment [23], $\theta_C$ of different signs and sizes were observed at different locations in a $CsV_3Sb_5$ sample. This can be explained if multi-domains of "sample 1" type and "sample 2" type are present in their sample, which will lead to location-dependent values of $\pm\theta_C$, and even intermediate values for sub-wavelength (1 $\mu m$) domains. It may also explain why their measured $\theta_C$ [23] is one order of magnitude smaller than possible "single domain" samples in this work and in another 800 $nm$ experiment where $\theta_C$ of the same sign and similar sizes were found in two $CsV_3Sb_5$ samples [24]. The sign of $\theta_C$ remains unchanged after thermal cycles in both sample 1 and 2, which agrees with the findings in the above 800 $nm$ experiment [23] that thermal cycles at the same location do not change $\theta_C$. In contrast, the MOKE signals $\theta_K$ (Fig.3(f)) flip sign with opposite fields and are identical to those measured in sample 1. At zero magnetic field the change of spontaneous $\theta_K$ across $T_{CDW}$ is below 0.02 $\mu rad$ (Fig.3(g)), which also agrees with the findings in sample 1.

With these observations, we conclude that the isotropic polarization component $\theta_C \sim \pm (1.3 \pm 0.1)\ mrad$ is not MOKE, which is five orders of magnitude smaller. And the sign of $\theta_C$, unlike the MOKE signal $\theta_K$, is unrelated to the direction of applied magnetic fields, but seems to have been predetermined at crystal growth. Therefore, $\theta_C$ is not related to the proposed TRSB loop currents[19,20] or experimentally observed TRSB signatures in both STM [15] and μSR [16–18] experiments. Being isotropic, $\theta_C$ also doesn't originate from anisotropy, which gives rise to the anisotropic component $\theta_P \sim 0.2\ mrad$.

What order does $\theta_C$ represent? Such an isotropic rotation at normal incidence may remind us of the controversial effect of specular (reflection) optical activity [38] in a chiral material such as cinnabar and cholesteric liquid crystals. Although this effect in normal incidence reflection is usually rejected [25], it was theoretically proposed to exist only in strongly absorbing materials and was reported in α-HgS (cinnabar) at 543 $nm$, close to the bandgap energy [38]. However, neither the 1550 $nm$ or the 800 $nm$ wavelength is close to a band gap in $CsV_3Sb_5$ [33,34]. In addition, as explained in [39], such specular optical activity is proportional to the difference of the magneto-electric tensor [37] components $k_{xx} - k_{yy}$ perpendicular to the propagation direction $z$, which is the c-axis of $CsV_3Sb_5$. As such, the resulting optical rotation would flip sign when the incidence polarization angle $\alpha$ is rotated by 90°, which is incompatible with the isotropic nature of $\theta_C$. In fact, this optical activity could produce an optical rotation $\propto (k_{xx} - k_{yy}) cos(2\alpha)$ (see discussions in the Supplementary Information) that would explain the puzzling sinusoidal form of the anisotropic rotation component $\theta_P sin(2\alpha - A)$.

Hence it is unlikely that isotropic $\theta_C$ originates from a chiral order such as a cholesteric CDW state. And we propose that this isotropic rotation $\theta_C$ observed at 800 $nm$ [23,24] and at 1550 $nm$ in this work represent a new intertwined order that generates an unconventional optical rotation not found in any other material system to our knowledge. Alternatively, it is possible that the complex intertwining of the various orders in the CDW state leads to novel hybrid phenomena such as the observed unconventional optical rotation. The topological Kagome lattice [40] is indeed full of surprises.

**Acknowledgements**: We acknowledge useful discussions with A. Kapitulnik. This project was supported by the Gordon and Betty Moore Foundation through Grant GBMF10276. S.D.W. and B.R.O. acknowledge support via the UC Santa Barbara NSF Quantum Foundry funded via the Q-AMASE-i program under award DMR-1906325.

**Author Contributions**: C.F. carried out the Sagnac MOKE measurements. J.W. performed the thermal measurements. S.D.W. and B.R.O. grew the crystals. J.X. carried out the optical rotation measurements and wrote the paper.

Supplementary Information for

# Unconventional Optical Rotation in the Charge Ordered State of Kagome Metal $CsV_3Sb_5$

Camron Farhang[1], Jingyuan Wang[1], Brenden R. Ortiz[2], Stephen D. Wilson[2] and Jing Xia[1]
[1]Department of Physics and Astronomy, University of California, Irvine, California 92697, USA
[2]Materials Department, University of California, Santa Barbara, Santa Barbara, CA 93106, USA.11973

## (A) Methods for measuring polarization rotation and MOKE

To resolve the controversy on the optical rotation of $CsV_3Sb_5$, it is important to measure both polarization rotation $\theta_T$ and MOKE $\theta_K$ on the same sample using the same optical wavelength and under the same experimental conditions. To achieve this, we have constructed a standard Wollaston-prism-based polarization rotation setup that can be connected to the same optical cryostat for the Sagnac interferometer microscope. This configuration allows both types of measurements on the same crystal (usually in the same region) without leaving the vacuum of the cryostat. Both instruments operate with continuous wave (CW) light sources at the 1550 $nm$ wavelength. The typical optical powers used in the experiments are 100 $\mu W$ in the polarization rotation setup and 20 $\mu W$ in the Sagnac interferometer. This amount of optical power has negligible heating effects at $T_{CDW} \sim 94\ K$. Normal incidence reflection is ensured in the polarization setup with a pinhole, and in the Sagnac interferometer with a single mode fiber.

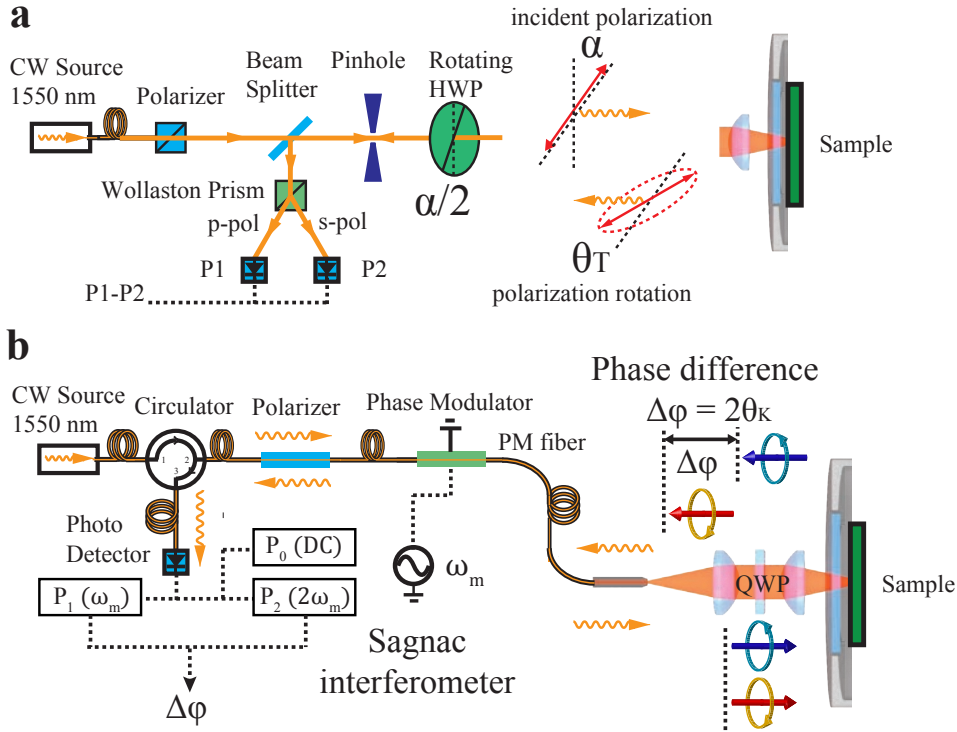

*Fig.S1. Polarization rotation and MOKE setups operating with continuous-wave (CW) light at 1550 nm wavelength:* Both optical setups are connected to the same optical cryostat to allow measurements on the same sample during one experiment. *(a)* Schematics of polarization rotation setup based on a Wollaston prism that measures polarization rotation $\theta_T$ as a function of incident polarization α, which is achieved by rotating a half-wave plate (HWP) by α/2. *(b)* Schematics of a zero-area-loop fiber-optic interferometer that is only sensitive to TRSB (MOKE $\theta_K$) effects, which is independent of α. The fiber-optic head can be scanned to simultaneously acquire reflection and MOKE images.

The schematics of the polarization rotation setup is shown in Fig.S1(a). The beam of light from a CW light source centered at 1550 $nm$ is routed through a free-space polarizer to produce a linearly polarized beam. A polarization-independent beam splitter (half mirror) transmits half of the beam and reflects the other half, which is discarded. The transmitted beam passes through a pinhole and then a half-wave plate (HWP), which is mechanically rotated such that its principle fast axis is at an angle $\alpha/2$ to the polarization direction of the beam. The resulting beam after the HWP has its polarization direction rotated by angle $\alpha$. The purpose is to allow the control of the relative angle $\alpha$ between the incident polarization and the same crystal axis without rotating the sample itself, which is difficult to do in the cryostat and will introduce noise and offsets. The beam then passes through the optical window of the cryostat and gets reflected by the sample. The returned light beam is in general elliptical with the major axis rotated by the total polarization rotation $\alpha + \theta_T$. After passing through the HWP a second time, its polarization direction is rotated by $-\alpha$, and becomes $\theta_T$. The optical path is aligned such that the beam can pass the same pinhole a second time, ensuring normal incidence reflection from the sample. Then the same polarization-independent beam splitter will reflect half the returned beam towards a Wollaston prism, which separates and directs two orthogonal polarizations s and p towards two balanced detectors. The recorded powers of s and p-polarization components are P1 and P2 respectively. The Wollaston prism is rotated at a $\pi/4$ angle such that with a gold mirror calibration sample ($\theta_T = 0$), P1 and P2 are "balanced": $\Delta P = P1 - P2 \sim 0$. In this configuration, one can show that the optical amplitudes $E1$ and $E2$ at detectors 1 and 2 are:

$$E1 = E0 \cos\left(\frac{\pi}{4} - \theta_T\right)$$

$$E2 = E0 \cos\left(\frac{\pi}{4} + \theta_T\right)$$

, where $E0$ is the total amplitude. Since optical intensity $I = E^2$, the sum and difference of the two intensities $I1$ and $I2$ are:

$$I1 + I2 = E1^2 + E2^2 = E0^2 \cos^2\left(\frac{\pi}{4} + \theta_T\right) + E0^2 \cos^2\left(\frac{\pi}{4} - \theta_T\right) = E0^2$$

$$I1 - I2 = E1^2 - E2^2 = E0^2 \cos^2\left(\frac{\pi}{4} - \theta_T\right) - E0^2 \cos^2\left(\frac{\pi}{4} + \theta_T\right) = E0^2 \sin(2\theta_T)$$

Hence:

$$\frac{I1 - I2}{I1 + I2} = \sin(2\theta_T)$$

As optical power is proportional to intensity $P \propto I$, we can extract $\theta_T$ as:

$$\theta_T = \frac{1}{2}\arcsin\left(\frac{I1 - I2}{I1 + I2}\right) = \frac{1}{2}\arcsin\left(\frac{P1 - P2}{P1 + P2}\right) = \frac{1}{2}\arcsin\left(\frac{\Delta P}{P1 + P2}\right)$$

, where $\Delta P$ is read from a direct output of the balanced detector. Optical components such as the focusing lens and cryostat optical window have $\mu rad$ to $mrad$ levels of optical birefringence due to residual strains. Fortunately, these contributions are independent of sample temperature since these optical components are outside of the cryostat and are at the fixed room temperature. In this work, a base line value at 120 $K$, $\theta_T(120\ K)$ is subtracted from $\theta_T$ to remove this background.

The schematics of the zero-loop Sagnac interferometer used in this work is shown in Fig.S1(b). The beam of light from a CW light source centered at 1550 $nm$ is routed by a fiber-circulator to a fiber-polarizer, which polarizes the beam. The circulator transmits light from port 1 to port 2 and from port 2 to port 3 with better than 30 dB isolation in the reverse directions. After the polarizer the polarization of the beam is at 45° to the axis of a fiber-coupled electro-optic modulator (EOM), which generates 4.6 MHz time-varying phase shifts $\phi_m \sin(\omega t)$, where the amplitude $\phi_m = 0.92$ rad between the two orthogonal polarizations that are then launched into the fast and slow axes of a polarization maintaining (PM) single mode fiber. Upon exiting the fiber, the two orthogonally polarized linearly polarized beams are converted into right- and left-circularly polarizations by a quarter-wave plate (QWP) and are then focused through the optical window of the cryostat onto the sample. After reflection from the sample and passing through the optical window, the same quarter-wave plate converts the reflected beams back into linear polarization with exchanged polarization axes. The two beams then pass through the PM fiber and EOM but with exchanged polarization modes in the fiber and the EOM. At this point, the two beams have gone through the same path but in opposite directions, except for a phase difference of $\Delta\varphi$ from reflection off the magnetic sample and another time-varying phase difference by the modulation of EOM. This nonreciprocal phase shift $\Delta\varphi$ between the two counterpropagating circularly polarized beams upon reflection from the sample is twice the Kerr rotation $\Delta\varphi = 2\theta_K$. The two beams are once again combined at the detector and interfere to produce an optical signal $P(t)$:

$$P(t) = \frac{1}{2}P[1 + \cos(\Delta\varphi + \phi_m \sin(\omega t))]$$

, where P is the returned power if the modulation by the EOM is turned off. For MOKE signals that are slower that the 4.590 MHz modulation frequency used in this experiment, we can treat $\Delta\varphi$ as a slowly time-varying quantity. And $P(t)$ can be further expanded into Fourier series with the first few orders listed below:

$$P(t)/P = \frac{1}{2}[1 + J_0(2\phi_m)]$$
$$+(\sin(\Delta\varphi)J_1(2\phi_m))\sin(\omega t)$$
$$+(\cos(\Delta\varphi)J_2(2\phi_m))\cos(2\omega t)$$
$$+2J_3(2\phi_m))\sin(3\omega t)$$
$$+\cdots$$

, where $J_1(2\phi_m)$ and $J_2(2\phi_m)$ are Bessel J-functions. Lock-in detection was used to measure the first three Fourier components: the average (DC) power (P0), the first harmonics (P1), and the second harmonics (P2). And the Kerr rotation can then be extracted using the following formula:

$$\theta_K = \frac{1}{2}\Delta\varphi = \frac{1}{2}tan^{-1}[\frac{J_2(2\phi_m)P1}{J_1(2\phi_m)P2}]$$

The noise in Kerr signal is shot-noise-limited to $10^{-7} rad/\sqrt{Hz}$ with 10 $\mu W$ of optical power, which is small enough not to heat up the sample even at the base temperature of the cryostat. By averaging over 100 seconds, 10 nanoradian (nrad) Kerr resolution can be achieved over a few Kelvins variation of sample temperatures. In practice, the bias offset in our system drifts about 20 nrad in experiments that take a long time or over wide sample temperature ranges. And the flexible fiber head can be mechanically scanned to simultaneously produce reflection (P0) and MOKE ($\theta_K$) images.

## (B) Cross-checking both optical setups with test samples

We have used a few test samples to cross-check the polarization rotation setup and the Sagnac interferometer. They are an uncoated gold mirror to evaluate offsets and noise, a magneto-optic thin film with a known MOKE signal to double check their calibrations, and a birefringent polyethylene polymer film to demonstrate the two setups different response to TRSB and non-TRSB optical rotations. Both total polarization rotation $\theta_T$ and the MOKE $\theta_K$ are measured as a function of the relative angle $\alpha$ between the incident polarization and the sample. With the polarization rotation setup, angle $\alpha$ is achieved by rotating the half-wave plate (HWP) by an angle $\alpha/2$. With the Sagnac interferometer, it is achieved by mounting the test samples on a rotational stage, which is rotated by angle $\alpha$.

The results on the gold mirror are plotted in Fig.S2(a), with the polar plots shown in Fig.S1(b). The uncoated gold mirror should introduce near-zero optical rotations and thus serves as a null test sample. The MOKE $\theta_K$ readings (red squares) are zero with 10s of $nrad$ uncertainty, regardless of the sample angle $\alpha$. This is the expected behavior for a Sagnac interferometer on a non-TRSB sample. The measured total polarization rotation $\theta_T$ (blue circles) scatter between 0 $\mu rad$ and 30 $\mu rad$ with most of the points falling near 20 $\mu rad$. The 20 $\mu rad$ is the offset mostly likely due to slight misalignments of the Wollaston prism, which will be eliminated in temperature-dependent measurements of CsV$_3$Sb$_5$ by subtracting the base line value at 120 $K$ sample temperature, $\theta_T(120\ K)$. The 30 $\mu rad$ scattering represents the instrumentation noise due to various effects such as vibration and air flow and can be much larger in the CsV$_3$Sb$_5$ measurements due to the introduction of the cryostat. We note that these noise sources are non-TRSB and are rejected by the Sagnac interferometer.

To ensure that the scale calibrations of both instruments are correct, we measure a magneto-optic (MO) film (XP33BC32) that generates a MOKE signal of around 4 $mrad$ at the 1550 $nm$ wavelength. Such a MOKE signal would present itself as an isotropic (polarization independent) rotation component $\theta_T(\alpha) = \theta_C$ in the polarization rotation measurement. A small magnetic field of 5 $mT$ is applied to align the ferromagnetic domains in the MO film. The results and their polar plots are shown in Fig.S2(c) and Fig.S2(d). The measured polarization rotation $\theta_T(\alpha)$ (blue circles) vary between 4.38 $mrad$ and 4.43 $mrad$ between incident polarization angles $\alpha$, and can be regarded as mostly isotropic, which is expected for a MOKE signal. The 50 $\mu rad$ or 1% variance could arise from small linear birefringence in the film due to residual strains but could also be due to the instrumental noise and drift. The measured MOKE signal $\theta_K(\alpha)$ (red squares) vary between 4.35 $mrad$ and 4.47 $mrad$ between sample angles $\alpha$. This 80 $\mu rad$ (2 %) variance is much larger than the 0.02 $\mu rad$ noise level of a Sagnac interferometer, whose reading should also not depend on the sample angles $\alpha$. Since the sample is rotated mechanically in the Sagnac measurements, where the axis of rotation is not necessarily aligned with the optical beam, we suspect that the optical beam was probing different locations along a circular path at different sample angles $\alpha$. And the inhomogeneity in the sample is the source of this 80 $\mu rad$ variance. Nevertheless, this MO film demonstrates that the calibrations of the polarization setup and the Sagnac interferometer agree within 2 %, which is more than enough to compare the results from both instruments with sufficient accuracy and to reach the conclusions of this paper.

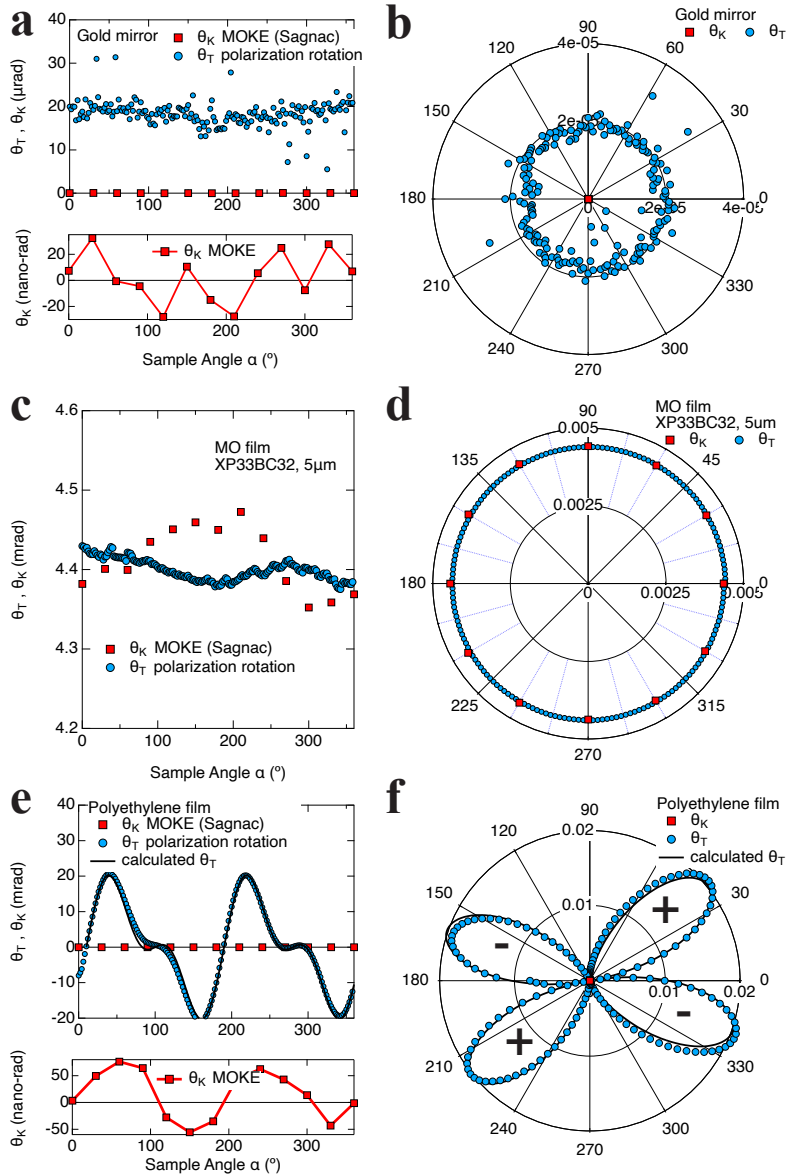

***Fig.S2. Cross-checking polarization rotation $\theta_T$ and MOKE $\theta_K$ on test samples.*** *Sample angle α is changed by sample rotation with the Sagnac MOKE setup, and by rotating the half-wave plate by α/2 in the polarization setup. **(a)(b)** An uncoated gold mirror with zero expected rotations. Measured $\theta_K$ is smaller than 30 nrad while $\theta_T$ has an instrument related 20 μrad offset. **(c)(d)** A magneto-optic film with ~ 4 mrad of MOKE signal. Measured $\theta_K$ and $\theta_T$ agree with each other within 2%. **(e)(f)** An anisotropic polyethylene film. Measured $\theta_K$ is smaller than 30 nrad as expected from a time-reversal symmetry invariant sample. $\theta_T$ shows a pattern that clearly demonstrates an anisotropic (rotational symmetry breaking) component $\theta_P$, with zero isotropic (polarization-independent) rotation component ($\theta_C$). Solid line is the calculated polarization rotation for an anisotropic reflective sample, and it agrees well with the measured $\theta_T$.*

Finally, a polyethylene film is used to demonstrate the different responses to different polarization components between the two instruments. This polymer film is optically anisotropic and thus produces anisotropic optical rotations. The results and their polar plots are shown in Fig.S2(e) and Fig.S2(f). The measured MOKE signal $\theta_K(\alpha)$ (red squares) remains zero as the sample doesn't break time-reversal symmetry, while the measured polarization rotation $\theta_T(\alpha)$ (blue circles) up to

$\pm 20$ $mrad$ displays a four-leaf clover shape with 2-fold rotational symmetry. The shape is a direct result of the presence of optical linear birefringence (LB) and optical linear dichroism (LD). And it can be calculated analytically (see section F) and be compared with the experimental curve to fit LB and LD parameters. In Fig.S2(e) and Fig.S2(f) I present the calculated $\theta_T(\alpha)$ curve (black line) using parameters LB = 0.133 and LD = 0.015, which matches well to the experimental data (blue circles). This polymer film serves as an example that while the polarization rotation setup detects the total rotation, the Sagnac interferometer is sensitive only to TRSB effects. We also note that in the reflection geometry, simple linearly birefringent and dichroic materials will often have a four-leaf clover shaped $\theta_T(\alpha)$ instead of a simple sinusoidal form $\theta_P \sin(2\alpha - A)$ found in AV$_3$Sb$_5$.

## (C) Full polarization rotation data of sample 1

In Fig.S3 we present polar plots of $\theta_T(\alpha)$ in CsV$_3$Sb$_5$ sample 1 at more temperatures. Fitted parameters of $\theta_C(T)$ and $\theta_P(T)$ are presented in the main text Fig.1(d).

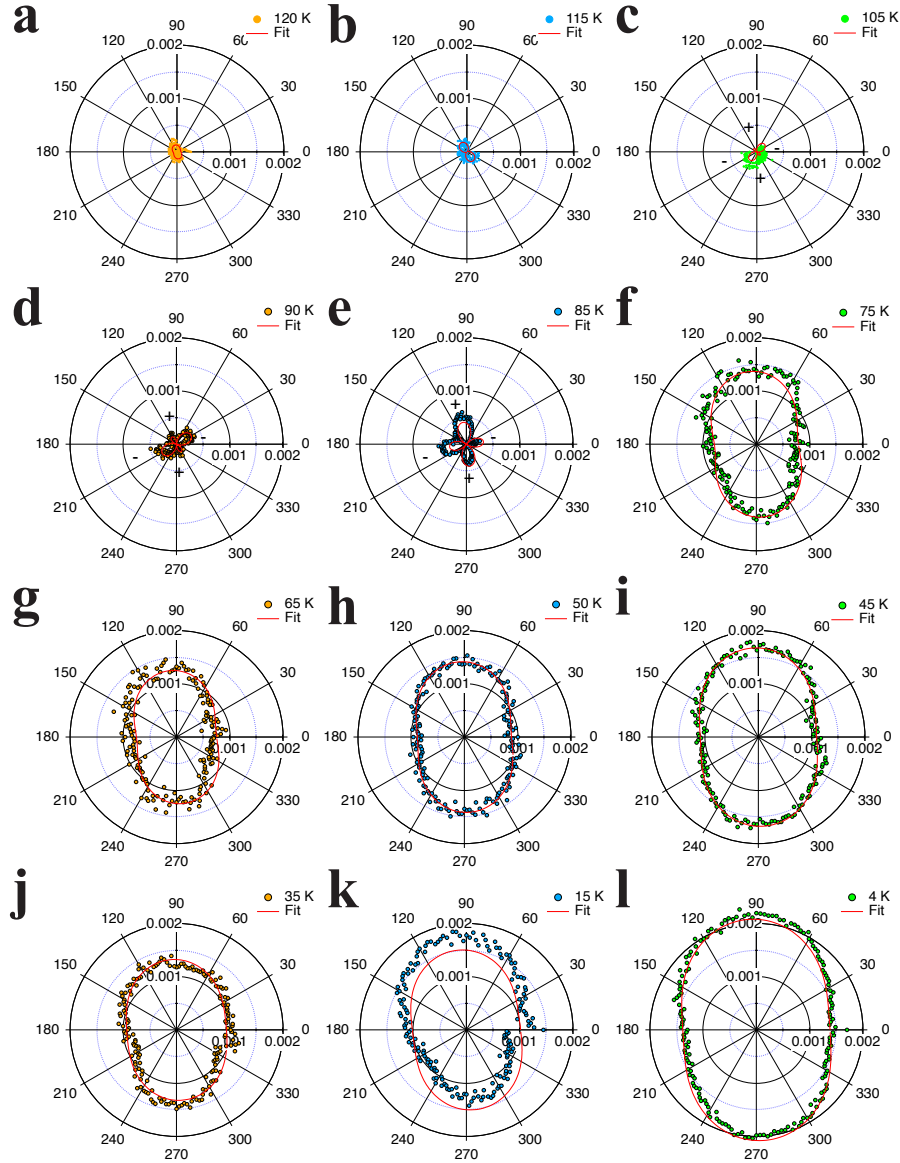

*Fig.S3. Polarization rotations in sample 1: (a)-(l) Polar plots of polarization rotations and fittings in sample 1 at various temperatures at zero magnetic field. The "-" sign indicates negative values.*

## (D) Full polarization rotation data of sample 2

In Fig.S4 we present polar plots of $\theta_T(\alpha)$ in CsV$_3$Sb$_5$ sample 2. Fitted parameters of $\theta_C(T)$ and $\theta_P(T)$ are presented in the main text Fig.3(b).

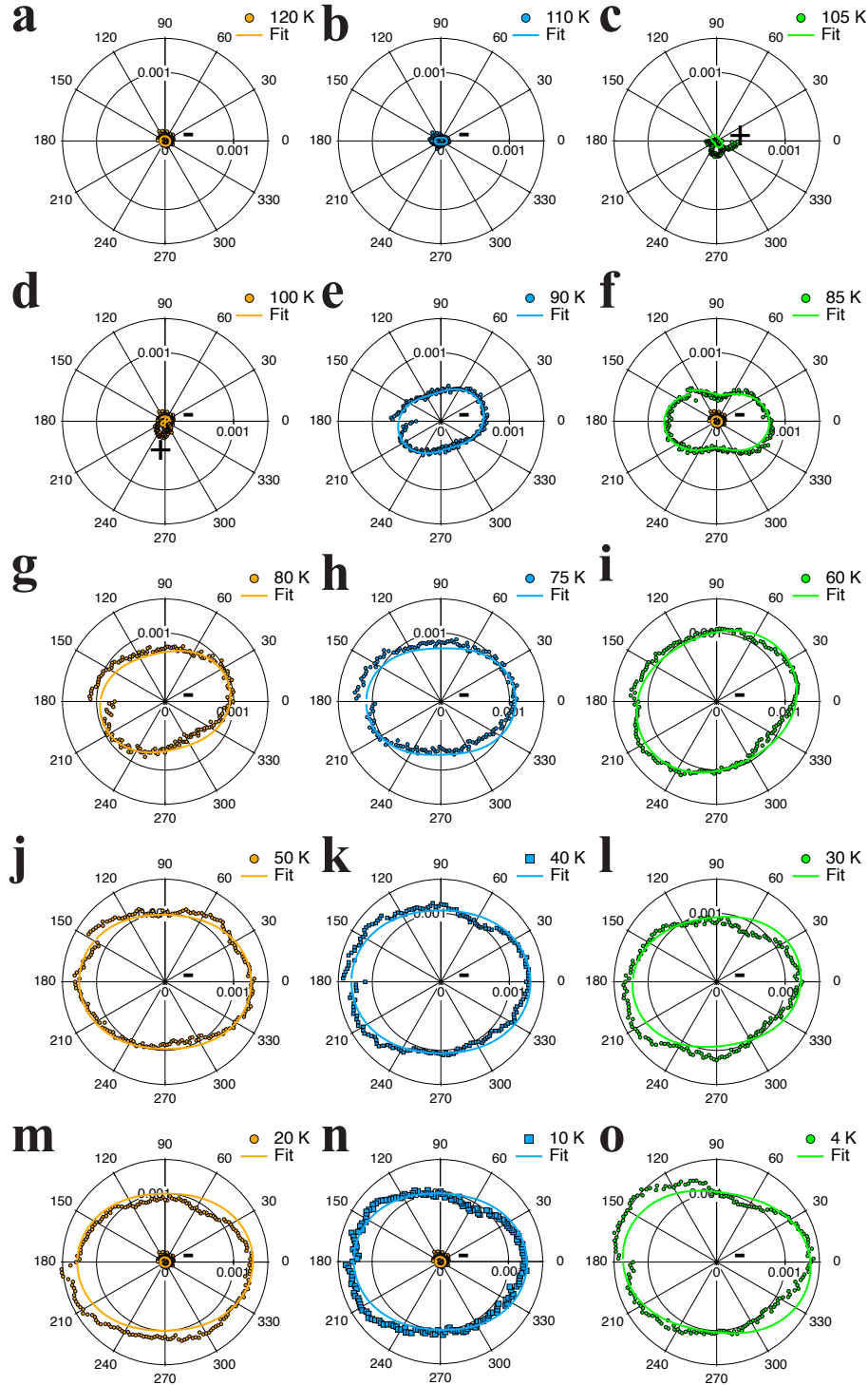

*Fig.S4. Polarization rotations in sample 2: (a)-(o)* Polar plots of polarization rotations and fittings in sample 2 at various temperatures with zero magnetic field. The "-" sign indicates negative values.

## (E) Full MOKE data of sample 2

In Fig.S5 we present temperature traces of $\theta_K(T)/B$ in CsV$_3$Sb$_5$ sample 2 with a magnetic field, and temperature traces of $\theta_K(T)$ during zero field warmup (ZFW). 1 μrad/T or 1 μrad offsets are introduced for clarify. These traces are numerical averaged to generate traces that are presented in the main text Fig.3(f) and (g).

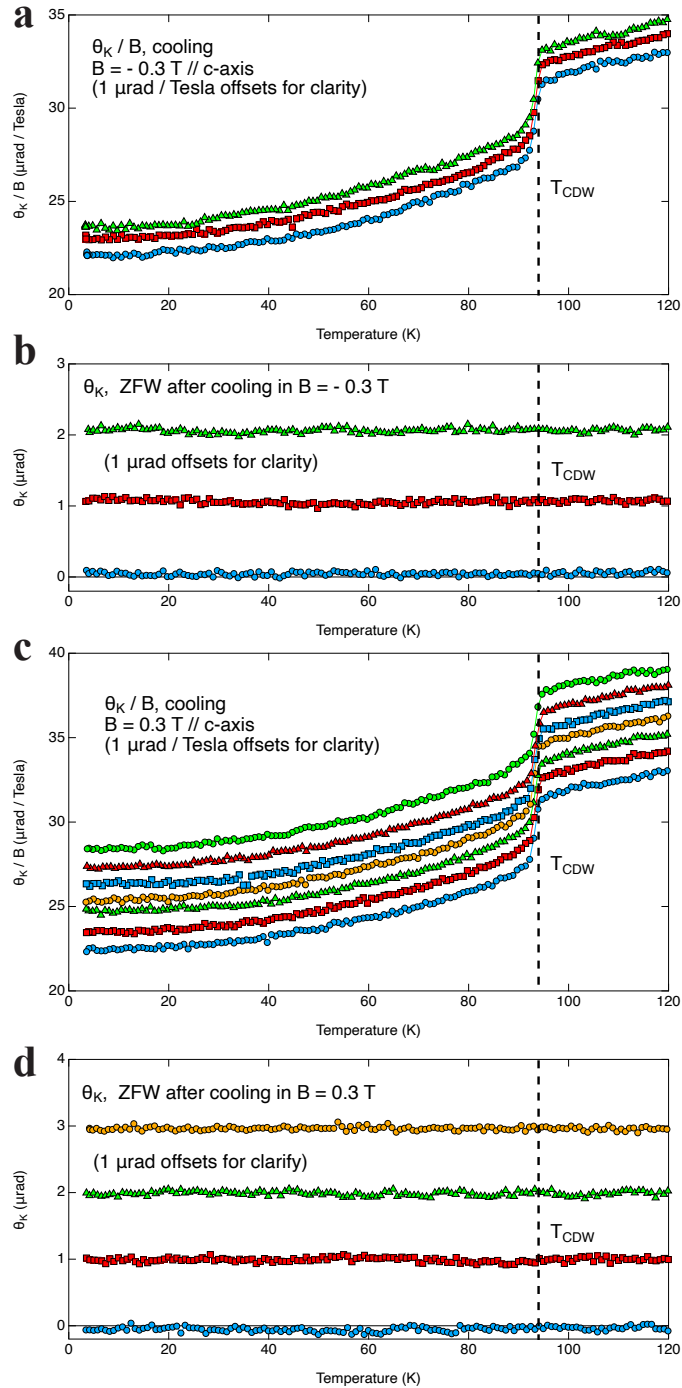

*Fig.S5. MOKE data of sample 2: (a) $\theta_K/B$ during cooldowns in -0.3 T magnetic field, with 1 μrad/T offsets for clarity. (b) $\theta_K$ during subsequent zero-field warmups (ZFW), with 1 μrad offsets for clarity. (c) $\theta_K/B$ during cooldowns in 0.3 T magnetic field, with 1 μrad/T offsets for clarity. (d) $\theta_K$ during subsequent zero-field warmups (ZFW), with 1 μrad offsets for clarity.*

### (F) The origin of the observed anisotropic rotation component $\theta_P \sin(2\alpha - A)$

The observed total polarization $\theta_T(\alpha)$ in both 800 $nm$ reflection experiments and in this work (1550 $nm$) contains an anisotropic component in the form of $\theta_P \sin(2\alpha - A)$, which has often been thought to be caused by optical linear anisotropy (linear birefringence and dichroism). However, using the Jones matrix, one can show that the polarization rotation of a sample with linear birefringence (LB) and linear dichroism (LD) has the following analytical form:

$$\theta_T(\alpha) = \sin(2\alpha)\cos(2\alpha) \frac{e^{i\,2\,LB}((LD-2)\,LD\,(2+(LD-2)\,LD)\,e^{i\,2\,LB}+(-(LD-1)^2-e^{i\,4\,LB}\,(LD-1)^2 +e^{i\,2\,LB}(2+(LD-2)\,LD\,(2+(LD-2)\,LD))))}{2+(LD-2)\,LD\,(2+(LD-2)\,LD)+(-2+LD)\,LD\,(2+(LD-2)\,LD)\cos(2\alpha))}$$

, which usually has the rather complicated four-leaf clover shape in Fig.S2(e) and Fig.S2(f). If linear dichroism LD is not present, the above formula is reduced to a simpler form:

$$\theta_T(\alpha) = \sin^2(LB)\sin(4\alpha)$$

It displays a 4-fold rotational symmetry pattern with sample angle $\alpha$, which is still different from the observed 2-fold symmetric pattern $\theta_P \sin(2\alpha - A)$ in CsV$_3$Sb$_5$ experiments. Therefore, the observed anisotropic $\theta_P$ component can't be explained by simple linear birefringence and/or dichroism that are often associated with a nematic CDW.

An alternative explanation is the reflection optical activity. As explained in *Opt. Lett.* **40**, 4277 (2015), at normal incidence reflection, such reflection optical activity is proportional to the difference of the magneto-electric tensor (*Rev. Mod. Phys.* **9**, 432–457 (1937)) components $k_{xx} - k_{yy}$ perpendicular to the propagation direction z, which is the c-axis of CsV$_3$Sb$_5$ in this work. Here the magneto-electric tensor describes the light-induced magnetization. As such, the resulting optical rotation would flip sign when the incidence polarization angle $\alpha$ is rotated by 90°, which is incompatible with the isotropic rotation component $\theta_C$, but could explain the anisotropic rotation component $\theta_P \sin(2\alpha - A)$ as the following. We consider the sub-tensor $K_{2D}$ of the magneto-electric tensor in the $xy$-plane perpendicular to the incident light:

$$K_{2D} = \begin{bmatrix} k_{xx} & k_{xy} \\ -k_{xy} & k_{yy} \end{bmatrix}$$

Under a sample rotation $\alpha$, this sub-tensor is transformed into:

$$K_{2D}(\alpha) = \begin{bmatrix} \frac{1}{2}(k_{xx}+k_{yy}+(k_{xx}-k_{yy})\cos(2\alpha)) & k_{xy}+\frac{1}{2}(k_{xx}-k_{yy})\sin(2\alpha) \\ -k_{xy}+\frac{1}{2}(k_{xx}-k_{yy})\sin(2\alpha) & \frac{1}{2}(k_{xx}+k_{yy}-(k_{xx}-k_{yy})\cos(2\alpha)) \end{bmatrix}$$

And the polarization rotation due to reflection optical activity is proportional to the difference between its diagonal components:

$$\theta_{OA}(\alpha) \propto K_{2D}(\alpha)_{xx} - K_{2D}(\alpha)_{yy} = (k_{xx} - k_{yy})\cos(2\alpha)$$

, which matches the observed sinusoidal form of the anisotropic rotation component $\theta_P \sin(2\alpha - A)$, if angle A is set to $\pi/2$. Therefore, we propose that an anisotropic order in the CDW state generates a difference between $k_{xx}$ and $k_{yy}$ in the magneto-electric tensor, where $xy$ plane is parallel to the cleaved surface perpendicular to the c-axis. It leads to an optical rotation component $\theta_P \sin(2\alpha - A)$ that is observed in both 800 $nm$ reflection experiments and in this work (1550 $nm$).